\documentclass[aps,pre,showpacs,twocolumn,floats]{revtex4}
\usepackage{amssymb}

\usepackage{amsmath}
\usepackage{graphicx,psfig}
\usepackage{dcolumn}
\usepackage{bm}

\input{tcilatex}

\begin{document}

\title{Quantum Dynamics of a Nanomagnet in a Rotating Field.}
\author{ C. Calero$^1$, E. M. Chudnovsky$^1$, and D. A. Garanin$^2$}
\affiliation{ \mbox{$^1$Physics Department, Lehman College, City
University of New York,} \\ \mbox{250 Bedford
Park Boulevard West, Bronx, New York 10468-1589, U.S.A.} \\
\mbox{$^2$Institut f\"ur Physik, Johannes-Gutenberg-Universit\"at,
 D-55099 Mainz, Germany}}
\date{\today}

\begin{abstract}
Quantum dynamics of a two-state spin system in a rotating magnetic
field has been studied. Analytical and numerical results for the
transition probability have been obtained along the lines of the
Landau-Zener-Stueckelberg theory. The effect of various kinds of
noise on the evolution of the system has been analyzed.

\end{abstract}
\pacs{75.50.Xx, 03.65.Xp}

\maketitle
\section{Introduction}

Molecular magnets with high spin and high magnetic anisotropy can
be prepared in long-living excited quantum states by simply
applying a magnetic field \cite{Sessoli}. In a time dependent
magnetic field, they exhibit stepwise magnetic hysteresis due to
resonant quantum tunneling between spin levels \cite{Friedman}.
This phenomenon has been intensively studied theoretically within
models employing Landau-Zener-Stueckelberg (LZS) effect
\cite{Landau,Zener,Stueckelberg}. The formulation of the problem,
independently studied by Landau \cite{Landau}, Zener \cite{Zener},
and Stueckelberg \cite{Stueckelberg} at the inception of quantum
theory is this. Consider a system characterized by quantum states
$|1\rangle$ and $|2\rangle$ with energies $E_1$ and $E_2$
respectively. Let the system be initially prepared in the
lower-energy state, $|1\rangle$, and the field be changing such
(due to, e.g., Zeeman interaction of the magnetic field with a
spin) that $E_1$ is shifting up while $E_2$ is shifting down.
After the levels cross and the distance between them continues to
increase, the system, with LZS probability, $P = \exp (-\pi
\Delta^2/2\hbar v)$, remains in the state $|1\rangle$. Here
${\Delta}$ is the tunnel splitting of $|1\rangle$ and $|2\rangle$
at the crossing, and $v$ is the rate at which the energy bias
between $|1\rangle$ and $|2\rangle$ is changing with time. This
picture, of course, does not take into account any disturbance of
the quantum states, $|1\rangle$ and $|2\rangle$, by the
dissipative environment. Application of the conventional LZS
effect to molecular magnets was initially suggested in Refs.
\onlinecite{Dobrovitski,Gunther}. Its dissipative counterpart was
developed in Refs. \onlinecite{Pokrovsky1, Pokrovsky2, Pokrovsky3,
Loss, Sinitsyn, Chudnovsky, Garanin4, Kayanuma, Kayanuma2}. The
amazing property of the LZS formula is that it is very robust
against any effect of the environment \cite{Sinitsyn2, Loss,
Chudnovsky}. This has allowed experimentalists to use the LZS
expression to extract ${\Delta}$ in molecular magnets from bulk
magnetization measurements
\cite{Wernsdorfer, Wernsdorfer2, Sarachik, Kent}.\\

The purpose of this paper is to elucidate the possibility of a
detailed study of quantum spin transitions and the effect of the
environment in experiments with a rotating magnetic field. Quantum
tunneling rates for a spin system in a rotating field have been
studied before \cite{Leo}. Here we are taking a different angle at
this problem, by computing the occupation numbers for quantum spin
states. This approach can be useful for the description of
experiments that measure the time dependence of the magnetization.
A weak high-frequency rotating field, $H_0 \sim 1\,$Oe, can be
easily achieved electronically, by applying $H_x = H_0\cos(\omega
t)$ and $H_z = H_0 \sin (\omega t)$. The rotating field of large
amplitude can be achieved by rotating the sample in a constant
magnetic field. In a typical molecular magnet, a rotating field
not exceeding a few kOe will result in the crossing of two spin
levels only, preserving the two-state approximation. We will
compute the time evolution of the probability to occupy one of the
two spin levels after a number of revolutions. We will demonstrate
that this evolution depends crucially on whether the system is
subject to the dissipative noise and that it depends strongly on
the frequency of the noise. The equivalent of the LZS effect in
the rotating field is studied in Sec. III. The effect of slow and
fast noise on such probability in the rotating field is considered
in Sec. IV . Consequences for experiment are discussed at the end
of the paper.

\section{Hamiltonian}

We shall start with the Hamiltonian
\begin{equation}\label{biaxial}
{\cal{H}} = -DS_z^2 - g{\mu}_B {\bf H}\cdot {\bf S}\, ,
\end{equation}
where $D$ is the uniaxial anisotropy constant, $g$ is the
gyromagnetic factor, ${\mu}_B$ is the Bohr magneton, and ${\bf H}$
is the magnetic field, and $S$ is an integer spin. We shall assume
that the magnetic field rotates in the $XZ$-plane,
\begin{equation}\label{magnetic field}
{\bf H} = H \sin(\omega t){\bf e}_z + H \cos(\omega t){\bf e}_x\,.
\end{equation}
At ${\bf H} = 0$, the ground state of the model is double
degenerate. The lowest energy states correspond to the parallel
($m = S$) and antiparallel ($m=-S$) orientation of ${\bf S}$ with
respect to the anisotropy axis, with $m$ being the magnetic
quantum number for ${\bf S}$. The effect of the external magnetic
field is twofold. The $Z$-component of the field removes the
degeneracy. The $X$-component produces a term in the Hamiltonian
that does not commute with $S_z$. Consequently, at $H_x \neq 0$
the $\left| m \right\rangle $ states are no longer the eigenstates
of the system. However, at
\begin{equation}\label{h-condition}
g{\mu}_B H \ll (2S-1)D
\end{equation}
we can treat the non-commuting term in the Hamiltonian as a
perturbation. Throughout this article it will be assumed that the
system is prepared initially in one of the saturated magnetic
states, say $\left|-S \right\rangle $ for certainty. This can be
easily achieved at low temperature in molecular magnets with high
easy-axis anisotropy.

For the perturbation $ V(t) = g{\mu}_B {\bf H}\cdot{\bf S}$ the
time-dependent perturbation theory gives the following expression
for the transition amplitude from the initial state, $\left| -S
\right\rangle $, to any $\left| m' \right\rangle $ state with $m'
\neq -S$:
\begin{eqnarray}\label{perturbation theory}
c_{m'}(t) & = & e^{-i\frac{E_{m'} t}{\hbar}}\bigg[
-\frac{i}{\hbar} \left\langle m'\right|g{\mu}_B H S_x\left|
-S\right\rangle \times \nonumber
\\
& & \int_{0 }^{t}dt_1 e^{i\frac{(E_{m'}-E_{-S})t_{1}}{\hbar}}
\cos(\omega t_1) + \left({\frac{-i}{\hbar}}\right)^2 \times \nonumber \\
 & & \sum_{m''
}\left\langle m'\right|g{\mu}_B H S_x\left|
m''\right\rangle\left\langle m''\right|g{\mu}_B H
S_x\left| -S\right\rangle \times \nonumber \\
& & \int_{0}^{t}dt_1
e^{i\frac{(E_{m'}-E_{m''})t_1}{\hbar}}\cos(\omega t_1) \times \nonumber \\
& &
\int_{0}^{t_{1}}dt_2e^{i\frac{(E_{m''}-E_{-S})t_2}{\hbar}}\cos(\omega
t_2)+... \bigg]\,,
\end{eqnarray}
where $E_m = -Dm^2$ are the eigenstates of ${\cal{H}}_0 =
-DS_z^2$. When $ \omega \ll ({E_{m'} - E_{-S}})/{\hbar}$ for all
$m'$ , then the perturbation can be treated adiabatically. This
requires the condition
\begin{equation}\label{requirement}
\hbar \omega \ll {(2S-1)D}\,,
\end{equation}
that will be used throughout this paper. In molecular magnets, $D$
is of the order of $1\,$K. Consequently, any $\omega$ at or below
GHz range satisfies Eq.\ (\ref{requirement}).

At $T = 0$, equations (\ref{h-condition}) and (\ref{requirement}),
and the initial condition, allow one to limit the consideration by
the two lowest states, $\left|S\right\rangle $ and
$\left|-S\right\rangle $, of the unperturbed Hamiltonian.
Time-independent perturbation theory based upon the condition
(\ref{requirement}) permits the usual reduction of the spin
Hamiltonian to the effective two-state Hamiltonian for the tunnel
split states originating from $m = S$ and $m= -S$,
\begin{equation}\label{two-state hamiltonian}
{{\cal{H}}_{eff}} = - \frac{1}{2} h_0 \sin(\tau)\sigma_z +
\frac{1}{2} \Delta (\tau) \sigma_x \,.
\end{equation}
Here $\sigma_z,_x$ are the Pauli matrices, $h_0 = 2Sg{\mu}_BH$ is
the amplitude of the energy bias, $\tau = \omega t$ is
dimensionless time, and $\Delta(\tau)$ is the tunnel splitting of
$\left|S\right\rangle $ and $\left|-S\right\rangle $ due to the
transverse field $H_x(\tau)$ \cite{Garanin},
\begin{equation}\label{splitting}
\Delta(\tau) = \Delta_0 \cos^{2S}(\tau)\,,
\end{equation}
where
\begin{equation}\label{splitting-0}
\Delta_0 = \frac{8S^2D}{(2S)!}\left(\frac{h_0}{4DS}\right)^{2S}\,.
\end{equation}

An important observation that follows from Eq.\ (\ref{splitting})
is that at $\tau_n = (2n + 1) \pi/2$ (with $n = 0, \pm 1, \pm 2,
...$) the splitting is exactly zero. According to Eq.\
(\ref{splitting}), at large $S$, it decreases very fast as one
moves away from the level crossing, that occurs at $\tau_n = n
\pi/2$. This guarantees that the transitions between the two
states are localized in time at the level crossing. This is
clearly seen in Figure 1 that shows the effect of the perturbation
on the energy levels.
\begin{figure}
\unitlength1cm
\begin{picture}(11,5.5)
\centerline{\psfig{file=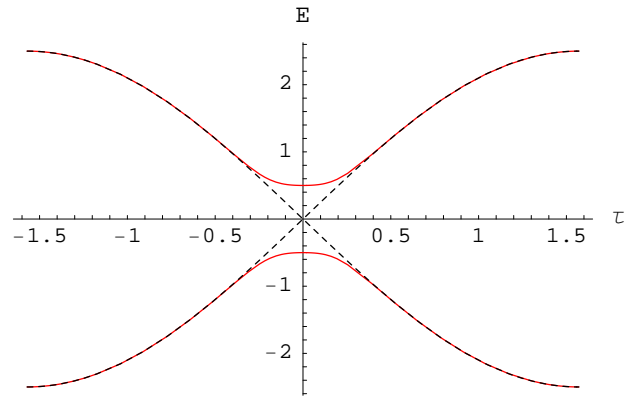,width=8.7cm}}
\end{picture}
\caption{\label{levels} Time dependence of the energy levels
(normalized by $D$) of the Hamiltonian (\ref{two-state
hamiltonian}) (solid line) at $S=4$ and $g{\mu}_B H/D = 3.1$. Dash
line shows the distance between the energy levels if they were
unperturbed by the second term in Eq.\  (\ref{two-state
hamiltonian}).}
\end{figure}

\section{Dynamics without noise}
\subsection{A single revolution}

In order to describe the evolution of the magnetization in a
magnetic field rotating in the $XZ$-plane, we shall compute first
the probability of staying at the initial state after a rotation
by 180 degrees, when a single level crossing takes place.

The Schr\"odinger equation for the coefficients of the wave
function,
\begin{equation}\label{Psi-c}
\left| \Psi\right\rangle = c_{-S}\left|-S\right\rangle + c_S\left|
S\right\rangle\,,
\end{equation}
can be expressed as:
\begin{eqnarray}\label{Schrodinger equation 1}
    & & {i\hbar} \frac{d}{dt}\tilde{c}_{-S}(t)  =  \frac{\Delta(t)}{2}  \tilde{c}_S(t) \nonumber\\
    & & {i\hbar} \frac{d}{dt}\tilde{c}_S(t)  =  -h_0 \sin(\omega t)\tilde{c}_S(t)+
    \frac{\Delta(t)}{2}
    \tilde{c}_{-S}(t)
\end{eqnarray}
where $ \tilde{c}_i(t) = c_i(t)\exp[{\frac{i}{2\hbar}
\int_{t_0}^{t}d{t'}h_0 \sin(\omega {t'})}]$. In terms of the
dimensionless variable,
\begin{equation}\label{u}
u = \frac{h_0\tau}{\Delta_0},
\end{equation}
Eq.\ (\ref{Schrodinger equation 1}) turns into:
\begin{eqnarray}\label{Schrodinger equation 2}
    & & i  \frac{d}{du}\tilde{c}_{-S}(u) =  \frac{{\tilde{\epsilon}}}{2} \cos^{2S}(\gamma u) \tilde{c}_S(u)
    \nonumber \\
    & & i  \frac{d}{du}\tilde{c}_S(u)   =  -\frac{\tilde{\epsilon}}{\gamma} \sin(\gamma u)\tilde{c}_S(u)+
    \frac{{\tilde{\epsilon}}}{2} \cos^{2S}(\gamma
    u)\tilde{c}_{-S}(u)\nonumber \\
\end{eqnarray}

The problem is now defined by two dimensionless parameters:
\begin{equation}\label{parameter-epsilon}
\tilde{\epsilon}  =  \frac{{\Delta_0}^2}{{\hbar\omega}\, h_0}\,,
\end{equation}
which is similar to the parameter used in the LZS theory
\cite{Garanin2, Dobrovitski}, and
\begin{equation}\label{parameter-gamma}
\gamma = \frac{\Delta_0}{h_0}\,,
\end{equation}
which is a measure of the magnitude of the magnetic field. Notice
that this parameter equals $\tau_c = \omega t_c$, where $t_c =
\Delta_0/(\omega h_0)$ is the characteristic time of crossing the
resonance. Thus, the condition $\gamma \ll 1$ is needed for the
crossing to be well localized in time on the time scale of one
revolution. This condition is required for the self-consistency of
the method.

From Eqs. (\ref{Schrodinger equation 2}) one can compute
numerically the time evolution of the coefficients $c_{-S}$ and
$c_S$, and, thus, the time evolution of the occupation numbers for
any $\tilde{\epsilon},\gamma$. In the fast ($\tilde{\epsilon} \ll
1 $) and slow ($\tilde{\epsilon} \gg 1)$ rotation regimes,
analytical formulas for the occupation probabilities can be
obtained. These formulas are useful for further analysis.
Inspection of Eqs.(\ref{Schrodinger equation 2}) reveals that the
deviation from the LZS result for slow rotation is small, since
within the relevant time of the transition $\delta u\sim 1$ and
$\Delta(\tau)$ is nearly constant. On the contrary, for the fast
rotation regime, ($\tilde{\epsilon} \ll 1$), the relevant time
interval is wider, $\delta u\sim \tilde{\epsilon}^{-1/2}$. This
allows a significant change of $\Delta$ during the transition and
makes possible a considerable deviation from the LZS result.

\subsubsection{Fast rotation ($\tilde{\epsilon} \ll 1 $)}
Following the procedure devised by Garanin and Schilling
\cite{Garanin2}, we can obtain the probability of staying at the
initial state after a 180-degree rotation of the external magnetic
field. We choose the direction of the magnetic field to be
initially antiparallel to the $Z$-axis. In the zero-th order of
the perturbation theory $\tilde{c_1}(u) = 1 $. In the first order,
such probability is then given by
\begin{equation}\label{Transition probability fast sweep}
  P =  1-\frac{1}{4} \left| \frac{\tilde{\epsilon}}{\gamma}
  \int_{-\frac{\pi}{2}}^{\frac{\pi}{2}} \cos^{2S}(z)
  \exp\left[{i\frac{\tilde{\epsilon}}{\gamma^2}\cos(z)}\right]dz \right|^2
\end{equation}
It is clear from this expression that for $\gamma \ll 1$, only
$|z| \ll 1$ contribute to the integral. This is in accordance with
the fact that the transition takes place during the time interval
${\Delta_0}/{h_0\omega}$, which is of order $\gamma$ compared with
the time of the integration. Consequently, one can approximate
$\cos(z)$ by
\begin{equation}\label{cos-approx}
\cos(z) \; {\approx} \; 1 -\frac{z^2}{2}\,, \qquad \cos^b(z) \;
{\approx} \; e^{-\frac{b}{2}z^2},
\end{equation}
where the exponential form is chosen to insure fast convergence of
the integral, and set infinite integration limits in Eq.\
(\ref{Transition probability fast sweep}):
\begin{eqnarray}\label{Transition probability fast sweep 2}
  P & = &1-\frac{1}{4} \left| \frac{\tilde{\epsilon}}{\gamma}
  \int_{-\infty}^{\infty}
  \exp\left[{-\left(\frac{i\tilde{\epsilon}}{\gamma^2} + S\right)z^2}\right]dz
  \right|^2 \nonumber \\
  & = & 1-\frac{\pi}{2} \frac{\tilde{\epsilon}}{\sqrt{1 +
  {({2S\gamma^2}/{\tilde{\epsilon}})}^2 }}\,.
\end{eqnarray}

\begin{figure}
\unitlength1cm
\begin{picture}(12,5.4)
\centerline{\psfig{file=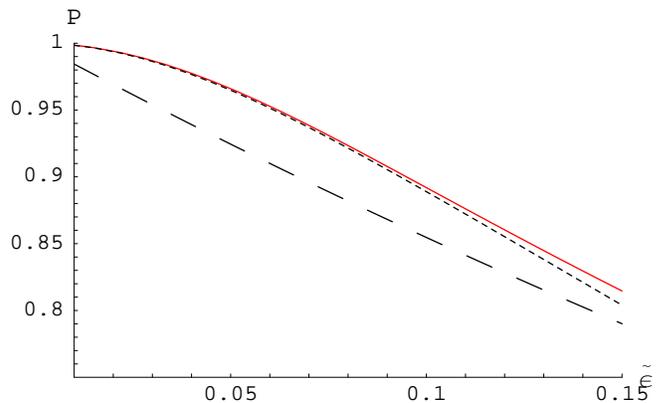,width=8.7cm}}
\end{picture}
\caption{\label{epsilon} Probability of staying at
$\left|-S\right\rangle$, $P$, as a function of parameter
$\tilde{\epsilon}$ at $\gamma = 0.1$\ and $S = 5$\,. The solid
line represents the numerical data, the short-dash line is the
analytical result, Eq.\ (\ref{Transition probability fast sweep
2}), and the long-dash line is the LZS result (see explanation in
the text).}
\end{figure}

\begin{figure}
\unitlength1cm
\begin{picture}(11,6)
\centerline{\psfig{file=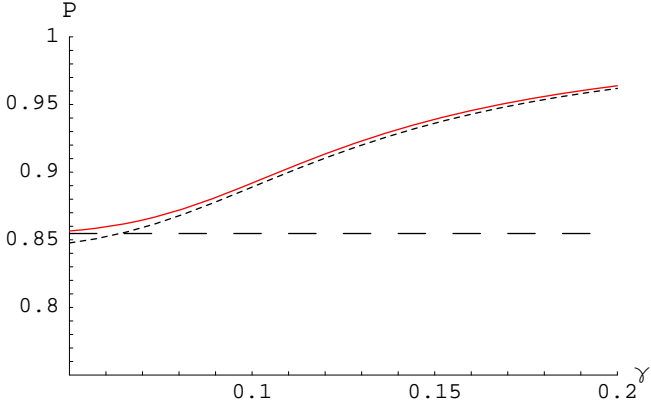,width=8.7cm}}
\end{picture}
\caption{\label{gamma} Probability of staying at the initial
state, $P$, as a function of $\gamma$ at $\tilde{\epsilon} = 0.1$
and $S = 5$\,. The solid line represents numerical data, the
short-dash line is the analytical result Eq.\,(\ref{Transition
probability fast sweep 2}), and the long-dash line is the LZS
result.}
\end{figure}

The probability, $P$, of staying at the initial state
$\left|-S\right\rangle$ after a 180-degree rotation is shown in
Fig.\ \ref{epsilon} for different values of $\tilde{\epsilon}$. In
the figure, the numerical results are compared with the result
given by Eq.\,(\ref{Transition probability fast sweep 2}), and
with the LZS result,
\begin{equation}\label{LZ}
P_{LZS} =\exp[{-{\pi\tilde{\epsilon}}/{2}}]\,.
\end{equation}
Note that Eq.\ (\ref{LZ}) would be our result for the probability
if we applied the LZS theory to the version of the Hamiltonian
(\ref{two-state hamiltonian}) that is linearized on $\tau$. As can
be seen from Fig.\ \ref{epsilon}, at $\gamma,\tilde{\epsilon} \ll
1$, Eq.\,(\ref{Transition probability fast sweep 2}) provides a
good approximation. The difference between the numerical result
and the LZS result is considerable. Fig.\ \ref{gamma} shows $P$
for different $\gamma$ at $\tilde{\epsilon} = 0.1$. Here again
Eq.\,(\ref{Transition probability fast sweep 2}), but not the LZS
formula, provides a good approximation for the $\gamma$-dependence
of the staying probability.

\subsubsection{Slow rotation ($\tilde{\epsilon} \gg 1 $)}
In the case of a slow rotation it is convenient to seek the
solution of the Schr\"odinger's equation in the adiabatic basis of
the two-state Hamiltonian. Then, one can follow a procedure
similar to that used for the fast rotation regime. The adiabatic
basis is given by:
\begin{equation}\label{Adiabatic basis}
\left|\Psi_{\pm}\right\rangle = \frac{1}{\sqrt{2}}(\pm
k_{\pm}\left|-S\right\rangle + k_{\mp}\left|S\right\rangle)
\end{equation}
where
\begin{equation}\label{k}
k_{\pm}(\tau) = \sqrt{1 \pm \frac{W(\tau)}{\sqrt{W^2(\tau) +
  \Delta^{2}(\tau)}}}\,.
 \end{equation}
with $W(\tau) = h_0 \sin(\tau)$ and $ \Delta(\tau) = \Delta_0
\cos^{2S}(\tau)$. The corresponding adiabatic energy levels are
\begin{equation}\label{Energy levels adiabatic basis}
  E_{\pm}(\tau) = \pm \frac{1}{2}\sqrt{W^2(\tau) +
  \Delta^{2}(\tau)}\,.
\end{equation}
Expressing the wave function as
\begin{equation}\label{wf}
\left| \Psi\right\rangle = c_+\left| \Psi_+\right\rangle +
c_-\left| \Psi_-\right\rangle
\end{equation}
we can now write the Schr\"odinger equation for
\begin{equation}
\tilde{c}_{\pm}(t) =
\exp\left[\frac{i}{\hbar}\int{dt{E_{-}(t)}}\right]c_{\pm}(t)
\end{equation}
in terms of the dimensionless variable u defined above:
\begin{eqnarray}\label{Sch eq. coefficients adiabatic base}
   & & \frac{d}{du}\,\tilde{c}_+ =-i\tilde{\epsilon}\Omega(\gamma
    u)\tilde{c}_+  - \frac{1}{2}\frac{d{w}/{du}}{(1+w^2)}\,\tilde{c}_- \nonumber \\
  & &  \frac{d}{du}\,\tilde{c}_- =
  \frac{1}{2}\frac{d{w}/{du}}{(1+w^2)}\,\tilde{c}_+\,,
\end{eqnarray}
where
\begin{equation}
\Omega(z) = \sqrt{\frac{1}{\gamma^2}\sin^2(z) + \cos^{4S}(z)}
\end{equation}
and $ w(z) = W(z)/\Delta(z)$.

\begin{figure}
\unitlength1cm
\begin{picture}(11,6)
\centerline{\psfig{file=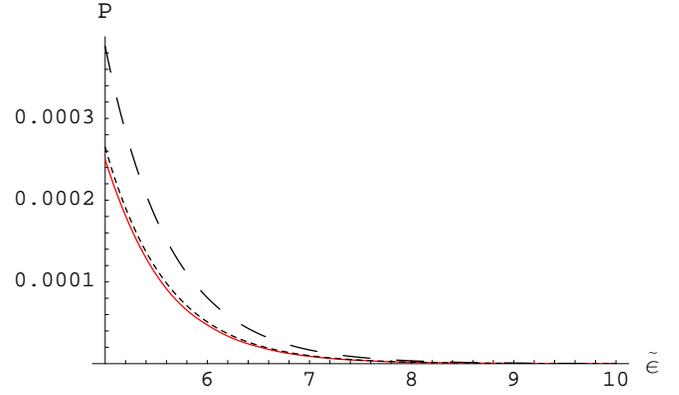,width=8.7cm}}
\end{picture}
\caption{\label{epsilonslow} Probability of staying at the initial
state, $P$, as a function of the parameter $\tilde{\epsilon}$ at
$\gamma = 0.1$ and $S = 5$\,. The solid line represents the
numerical data, the short-dash line is the analytical result, Eq.\
(\ref{Transition probability slow sweep2}), and the long-dash line
is the LZS result}
\end{figure}

The asymptotic behavior of $\tilde{c}_-(u)$ is $\tilde{c}_-(u)
\rightarrow 1$ for $\tilde{\epsilon} \rightarrow \infty$. At
$\tilde{\epsilon}\gg 1$ the coefficient $\tilde{c}_-(u)$ remains
close to 1. Also, far from the crossing point, $\tilde{c}_+(u)
\approx c_{-S}$, as can be seen from Eqs.\,(\ref{Adiabatic basis})
and (\ref{k}). These two facts allow one to obtain the probability
of staying at the initial state in the first order of perturbation
theory:
\begin{equation}\label{Transition probability slow sweep}
  P = \frac{1}{4}\left|
  \int_{-\frac{\pi}{2}}^{\frac{\pi}{2}}dz
  \frac{d{w}/{dz}}{(1+w^2)}\,
   \exp\left[{i\frac{\tilde{\epsilon}}{\gamma}\int_{0}^z dz'\,\Omega(z')}\right] \right|^2\,
\end{equation}
At $\gamma \ll 1$ the integral is dominated by $z$ close to zero.
The correct prefactor can be obtained by applying the procedure
outlined in Ref. \onlinecite{Garanin2}. With the accuracy to
$\gamma^2$ this gives:
\begin{equation}\label{Transition probability slow sweep2}
  P =
  \exp\left\{{-\frac{\pi\tilde{\epsilon}}{2}\left[1+\left(S-\frac{1}{8}\right)\gamma^2\right]}\right\}
\end{equation}
This probability is shown in Fig.\ \ref{epsilonslow} for different
values of $\tilde{\epsilon}$. The numerical results are compared
in the figure with the result given by Eq.\,(\ref{Transition
probability slow sweep2}), and with the LZS result. As can be seen
from Fig.\ \ref{epsilonslow}, at $\gamma \ll 1$,
Eq.\,(\ref{Transition probability slow sweep2}) provides a good
approximation for $\tilde{\epsilon} > 5 $.\\

\subsection{Continuous Rotation}

Our treatment of the continuous rotation is based upon the
smallness of the transition time in comparison with the period of
the rotation. Periodically driven two-state systems of that kind
have been studied before \cite{Kayanuma}.

The individual crossing is described by the transfer matrix:
\begin{equation}\label{matrix}
M = \left( \begin{array}{cl}
        \sqrt{P}  & e^{-i\theta}\sqrt{1-P}\, \\
        \\
        \ - e^{i\theta}\sqrt{1-P}\, &\;  \sqrt{P}
        \end{array} \right )
\end{equation}
where $\theta$ is the Stokes phase given by:
\begin{equation}\label{Stokes phase}
 \theta = \frac{\pi}{4} + \arg\left[\Gamma\left(1-i\frac{\tilde{\epsilon}}4\right)\right] +
\frac{\tilde{\epsilon}}{4}\left[\ln\frac{\tilde{\epsilon}}{4}
-1\right]\,.
\end{equation}
This matrix transforms a given initial state into the
after-crossing final state in terms of the unperturbed basis. The
above expression corresponds to the crossing in which the
$\left|-S\right\rangle$ level moves up towards the
$\left|S\right\rangle$ level that is moving down. In the opposite
case, $M$ should be replaced by the transpose matrix, $M^T$.

To describe the evolution of the system between crossings it must
be noted that, as we have shown before, far from the crossings the
unperturbed basis {$\left|-S\right\rangle , \left|S\right\rangle$}
almost coincides with the adiabatic basis {$\left|+\right\rangle ,
\left|-\right\rangle$}, see equations (\ref{Adiabatic basis}) and
(\ref{k}). In this region, the evolutions of
{$\left|-S\right\rangle$ and $\left|S\right\rangle$} are then
considered independent, so that they can be described by the
propagator
\begin{equation}\label{propagator}
G_n = \left( \begin{array}{cl}
            \exp[(-1)^{n+1}i\alpha] & 0 \\
            0 & \exp[(-1)^{n}i\alpha]
            \end{array} \right )
\end{equation}
where
\begin{equation}
\alpha =
\frac{1}{2}\left(\frac{D}{\hbar\omega}\right)\int_{0}^{\frac{\pi}{2}}d{\tau}\sqrt{W(\tau)^2
+ \Delta(\tau)^2}\,.
\end{equation}

\begin{figure}
\unitlength1cm
\begin{picture}(11,5.5)
\centerline{\psfig{file=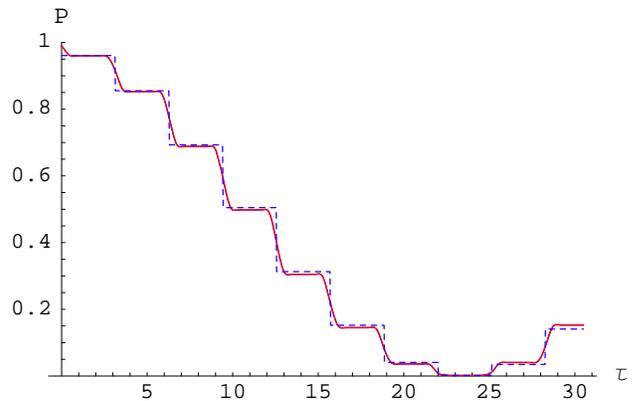,width=8.7cm}}
\end{picture}
\caption{\label{coherence} Time dependence of the probability of
finding a continuously rotating system in the initial state
$\left|-S\right\rangle$ for $\tilde{\epsilon} = 0.04, \gamma =
0.06$ and $S = 5$. The solid line represents numerical data. The
dash line shows analytical result obtained by successive
application of $M, M^T$ and $G_n$.}
\end{figure}

With the help of Eq.\ (\ref{matrix}) and Eq.\ (\ref{propagator})
one can compute the time evolution of the coefficients $c_{-S},
c_S$ in Eq.\ (\ref{Psi-c}). Starting with the initial state, the
state of the system after the n-th crossing can be obtained by the
successive action of $M, M^T$ and $G_n$. The time-dependence of
the probability of finding a continuously rotating system in the
initial state $\left|-S\right\rangle$ is shown in Fig.\
\ref{coherence}. The figure shows good agreement of the above
analytical method with numerical calculation. It is important to
notice that in the absence of dissipation the system does not
arrive to any asymptotic state at $\tau \rightarrow \infty$. The
behavior of the probability shows a long-term memory of the
initial state, which is somewhat surprising. This prediction of
the theory can be tested in real experiment.
\begin{figure}
\unitlength1cm
\begin{picture}(11,5.5)
\centerline{\psfig{file=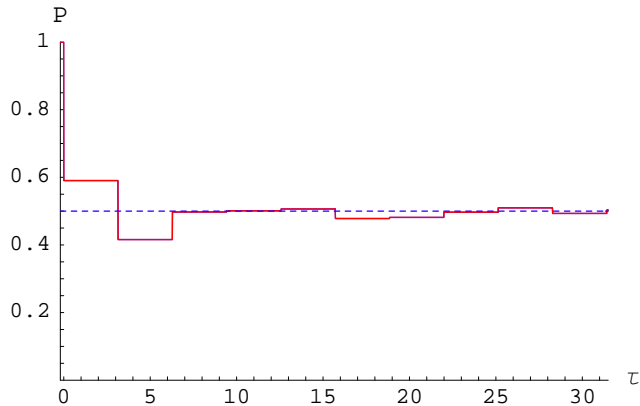,width=8.7cm}}
\end{picture}
\caption{\label{decoherence} Time dependence of the probability of
finding a particle in the state $\left|-S\right\rangle$ in the
presence of a low-frequency noise for $\tilde{\epsilon} = 1/3, S =
10$, and $\gamma = 0.01$. The plotted probability is the average
over an ensemble of 20 two-state dissipative systems.}
\end{figure}

\section{Dynamics with noise.}

When considering the effect of the noise, it is important to
distinguish between the following three regimes:
\begin{eqnarray}
& & \gamma \ll 1 \ll \omega/\Gamma \label{cond1} \\
& & \gamma \ll \omega/\Gamma \ll 1 \label{cond2} \\
& & \omega/{\Gamma} \ll \gamma \ll 1  \, \label{cond3}
\end{eqnarray}
where $\Gamma$ is the characteristic frequency of the noise. The
first of these conditions corresponds to the situation when a few
revolutions may occur before any contribution of the noise becomes
apparent. Consequently, during the time interval satisfying $t <
1/\Gamma$ one can use the results for the probability obtained in
the previous section. Under the condition (\ref{cond2}), one can
use the previously obtained results for a singular crossing but
needs to take into account the destruction of the relative phase
of the two states by the noise before the next crossing takes
place. Under the condition (\ref{cond3}) the results of the
previous section do not apply because the coherence of the quantum
state is destroyed by the noise on a timescale that is less than
the crossing time.

\subsection{Low-frequency noise, $\gamma \ll \omega/\Gamma \ll 1$}

The situation corresponding to the condition (\ref{cond2}) can be
easily described by the Hamiltonian
\begin{equation}\label{em-hamiltonian}
{\cal{H}}_{eff} = - \frac{1}{2} h_0 \sin(\tau)\sigma_z +
\frac{1}{2} \Delta (\tau) \sigma_x - \eta \sigma_z\,,
\end{equation}
where $\eta(\tau) \ll h_0$ is a random magnetic field in the
$Z$-direction, with the correlator
\begin{equation}\label{correlator}
\left\langle \eta(\tau) \eta(\tau')\right\rangle =
\eta_0^2\Theta\left[\left(\omega/\Gamma\right) -
\left|\tau-\tau'\right|\right]\,,
\end{equation}
$\Theta$ being the theta-function.

The time dependence of the probability of finding the system in
the state $\left|-S\right\rangle$ is shown in Fig.\
(\ref{decoherence}) for $\tilde{\epsilon} = 1/3, S = 10$, and
$\gamma = 0.01$. In accordance with the expectation, the
probability for the system to occupy the state with $m = -S$ (or
$m = S$), after going through a few oscillations, tends to the
asymptotic value of 0.5. For molecular magnets, only the average
of the probability over an ensemble of two-state dissipative
systems is of practical importance. The probability shown in Fig.\
(\ref{decoherence}) is such an average.

\subsection{High-frequency noise, $\omega/{\Gamma} \ll \gamma \ll 1$}

In this limit the coherence is completely suppressed, by, e.g.,
interaction with phonons, and the evolution of the population of
energy levels must be described by the density matrix. In this
case the population, $N_{-S}$, of the initially occupied state
$\left|-S\right\rangle$ is given by \cite{Garanin3, Chudnovsky}
\begin{equation}\label{incoherent evolution}
\frac{d{N}_{-S}}{dt} = -\frac{\Delta(t)^2}{2}
\frac{\hbar\Gamma}{W(t)^2 + (\hbar\Gamma)^2}(N_{-S}- N_{S})\,.
\end{equation}
The solution is
\begin{equation}\label{analytical solution incoherent case}
N_{-S} = \frac{1}{2}\left\{1 +
\exp\left[-\frac{\tilde{\epsilon}}{2}g\left(\tau;
2S;{\frac{\hbar\Gamma}{h_0}}\right)\right]\right\}\,,
\end{equation}
where $g(z; b;\alpha)$ is given by
\begin{eqnarray}\label{analytical solution incoherent case2}
& & g(z; b;\alpha) = \\
& &
(2n+2)f_1(\frac{\pi}{2})-f_1\left(\frac{\pi}{2}-\xi\right),\;\;
{\rm if} \;\;z = (2n+1)\frac{\pi}{2} + \xi
 \nonumber \\
& & g(z; b;\alpha) = 2n f_1(\frac{\pi}{2}) + f_1(\xi),\;\;{\rm if}
\;\; z = n\pi +\xi\,,
\end{eqnarray}
where $n = 0, 1, 2...$, $\;0 < \xi < \pi/2$,
\begin{equation}\label{analytical solution incoherent case3}
 f_1(z) = F_1\left(\frac{1}{2};
\frac{1}{2} - b, 1;\frac{3}{2}; \sin^2(z),
-\frac{\sin^2(z)}{\alpha^2}\right)\frac{\sin(z)}{\alpha}\;,
\end{equation}
and $F_{1}(a, b_1, b_2;c;x,y)$ is the Appell hypergeometric
function of two variables.

\begin{figure}
\unitlength1cm
\begin{picture}(11,5.5)
\centerline{\psfig{file=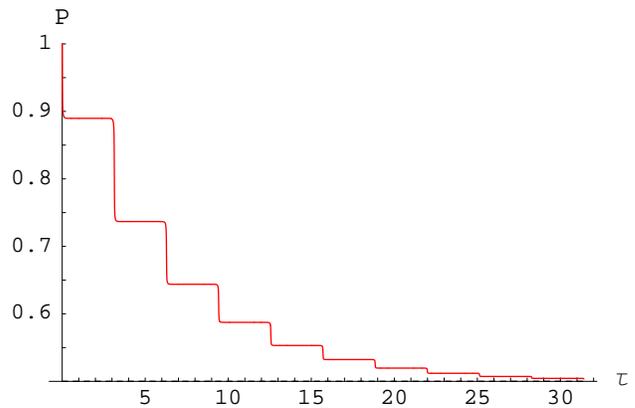,width=8.7cm}}
\end{picture}
\caption{\label{incoherence} Time dependence of the probability of
finding a particle in the state $\left|-S\right\rangle$ in the
presence of a high-frequency noise for $S = 10$, $\tilde{\epsilon}
= 1/3$ and $\frac{\Gamma}{h_0} = 0.01 $ .}
\end{figure}

The time-dependence of the occupation of the state
$\left|-S\right\rangle$ is shown in Fig.\ \ref{incoherence} for
$\tilde{\epsilon} = 1/3, S = 10, \hbar\Gamma/{h_0} = 0.01$. As in
the case of a low-frequency noise, the probability to occupy
either of the two levels tends to 0.5 after a few revolutions. The
difference between the two regimes is that in the case of the
high-frequency noise the probability monotonically approaches the
asymptotic value without exhibiting any oscillation.

\section{Conclusions}
We have studied the equivalent of the LZS effect for a spin system
in a rotating magnetic field. Typical time dependence of the
probability of staying at the initial state has been computed for
three different situations. The first is the situation when the
noise is irrelevant on the time scale of the measurement, Fig.\
\ref{coherence}. In this case the system exhibits coherent
behavior and long-term memory effects. The second situation
corresponds to the noise that decohere quantum states within the
time of each revolution but is slow enough to provide pure quantum
dynamics during the level crossing, Fig.\ \ref{decoherence}. The
third situation corresponds to a very fast noise that does not
allow the use of wave functions for the description of the
crossing and requires the density-matrix formalism, Fig.\
\ref{incoherence}. When the noise becomes important, the
occupation probability of each level approaches $1/2$ after
several revolutions. However, the asymptotic behavior depends on
the frequency of the noise. The three regimes discussed above are
given by equations (\ref{cond1}),(\ref{cond2}),(\ref{cond3}). One
must be able to switch between different regimes by changing the
angular velocity of the rotating field and/or temperature.
Experiments of that kind can shed light on the effect of
dissipative environment on the resonant spin tunneling in
molecular magnets. To be on a cautious side, one should notice
that the evolution of the magnetization in a crystal of magnetic
molecules also depends on the dipolar interactions between the
molecules \cite{Prokofev,Julio,Gar-Sch}. Our results are likely to
be relevant to molecular magnets when the amplitude of the
rotating field significantly exceeds dipolar fields. A candidate
for such a study would be, e.g., an uniaxial Ni-4 molecular magnet
that has no nuclear spins and that, with good accuracy, is
described by the Hamiltonian studied in this paper.

\section{Acknowledgements}
This work has been supported by the NSF Grant No. EIA-0310517.

\end{document}